\documentclass[preprint,12pt,3p]{elsarticle}

\usepackage{graphicx}
\usepackage{epstopdf,epsfig}
\usepackage{amssymb}
\usepackage{amsmath}
\usepackage{makecell}
\usepackage{tikz}
\usepackage{dirtytalk}
\usepackage{float}
\usepackage{tabularx}
\usepackage{adjustbox}
\usetikzlibrary{shapes}

\DeclareRobustCommand\sampleline[1]{%
\tikz\draw[#1] (0,0) (0,\the\dimexpr\fontdimen22\textfont2\relax)
-- (1.6em,\the\dimexpr\fontdimen22\textfont2\relax);}

\let\today\relax

\begin{document}
\begin{frontmatter}

\title{ Modeling the Dense Spray Regime Using an Euler-Lagrange Approach With Volumetric Displacement Effects}

\author[label1]{Pedram Pakseresht}
\address[label1]{School of Mechanical, Industrial and Manufacturing Engineering, Oregon State University, Corvallis, OR 97331, USA}

\author[label1]{Sourabh V. Apte\corref{cor1}}

\cortext[cor1]{Corresponding author. 204 Rogers Hall, Corvallis, OR 97331, USA. Tel: +1 541 737 7335, Fax: +1 541 737 2600. Email address: Sourabh.Apte@oregonstate.edu; pakserep@oregonstate.edu}

\begin{abstract}
Modeling of a dense spray regime using an Euler-Lagrange approach is challenging because of local high volume loading. A cluster of droplets, that are assumed subgrid, can lead to locally low void fractions for the fluid phase. Under these conditions, spatio-temporal changes in the fluid volume fractions should be considered in an Euler-Lagrange, two-way coupling model. This leads to zero-Mach number, variable density governing equations. Using pressure-based solvers, this gives rise to a source term in the pressure Poisson equation and a non-divergence free velocity field. To test the validity and predictive capability of such an approach, a round jet laden with particles is investigated using Direct Numerical Simulation coupled with point-Particle based model and compared with available experimental data for a particulate turbulent round jet with $Re_j=5712$. Standard force closures including drag, lift, Magnus effect, pressure, added mass as well as viscous torque acting on each individual particle are employed in the Point-Particle based model. In addition, volume displacement effects due to the presence of solid particles or liquid droplets, which is commonly neglected in the standard two-way coupling, are taken into account in both continuity and inter-phase momentum transfer to accurately capture the underlying structure of particle-turbulence interactions. Prediction results are in well agreement with the corresponding experiment. 
\end{abstract}

\end{frontmatter}

\section{Introduction}
\begin{figure}[t]
\setlength{\unitlength}{0.012500in}
\centering
\includegraphics[scale=0.5]{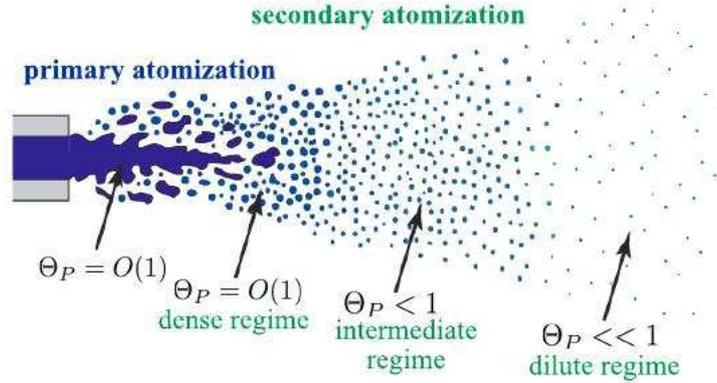}
\caption{Regimes of liquid spray evolution from injectors in gas-turbine engines}
\label{fig:spray}
\end{figure}

The performance of aircraft engines depends to a large extend on the efficiency and stability of the combustion process which in turn is extremely sensitive to the spatio-temporal distribution of the fuel/air mixture in the combustor. Fuel injection followed by atomization or breakup in the liquid fuel helps improve the combustion process with increasing the fuel surface area in order to decrease the time required for evaporating the fuel. The numerical models for spray calculations should be able to accurately represent droplet deformation, breakup, collision/coalescence, and dispersion due to turbulence. Significant progress has been made in modeling combustion in engines, yet accurate predictive models and simulation tools for dense spray modeling are still lacking. A standard modeling approach for liquid fuel atomization is to split the process into two subsequent steps: primary followed by secondary atomization shown in Fig.~\ref{fig:spray}. In this work, an Euler-lagrange approach taking into account the effect of finite size dispersed droplets are performed numerically to develop a technique for modeling the dense regime of sprays in the secondary atomization region. 

Concerning the combustion process and its efficiency, it is the secondary atomization region  where most phase change occurs, since there the liquid/gas surface area is orders of magnitude larger than in the primary atomization region, the gaseous temperature tends to be higher due to the closer proximity to the combustion zone, and the residence time of the liquid in the relatively large secondary atomization region is markedly longer than in the compact primary atomization region. The secondary atomization region is characterized by a vast number of droplets that interact with the surrounding gas transferring mass, momentum, and energy and can be characterized by three different regimes as shown in Fig.~\ref{fig:spray}. In the dense regime close to the primary atomization region, the liquid volume fraction $\theta_p$ is in the order of one with liquid droplets undergoing secondary breakup. In the intermediate regime, droplets continue to undergo further disintegration; however, $\theta_p$ is now smaller than unity. Finally, in the dilute regime, atomization is rare, $\theta_p$ is small, the droplets evaporate, and the fuel vapor mixes with the surrounding hot gases. The characterization of sprays based on physical processes in the different regimes also indicates different methodologies/models necessary to capture the dynamics of different regimes. For example, in the dense and intermediate regimes, not only are droplet deformation, collision, and coalescence important, but noticeable droplet-loading and variations in $\theta_p$ are crucial and should be 
taken into account to capture the spray evolution precisely. 

In the traditional approaches for spray modeling, the dynamics of the liquid/air interface are not resolved. In fact, the liquid phase is modeled through either an Eulerian (TFM) approach in which droplet considered to be as a continuous liquid phase or Lagrangian Point-Particle/Parcel method (PP) where droplets are assumed subgrid and their motion is captured by laws for drag, buoyancy and pressure forces, etc. from the gas phase. The effect of the droplets on the gas phase is modeled through two-way coupling of mass, momentum, and energy exchange \cite{Dukowicz1980}. Unlike most of the traditional models in which the liquid phase is assumed to break into a body of droplets as soon as it enters the combustion chamber, i.e. no primary atomization
(well-known as “Blob” hypothesis), the Eulerian-Lagrangian Spray Atomization (ELSA) approach couples the Eulerian mixing description for primary atomization (LES/DNS) with Lagrangian formulation (PP method) for secondary breakup \cite{Blokkeel2003,Lebas2005}. These models were originally derived in the context of RANS turbulence models and assume infinite Weber number; however, extensions to LES formulations have been recently proposed by \cite{Chesnel2011a,Chesnel2011b}. 

Recently, hybrid approaches of DNS method for the primary atomization region 
and LES (or DNS) coupled with Lagrangian Point-Particle/Parcel approach for solving the gas and liquid phases respectively in the secondary atomization region have been developed by \cite{Herrmann2010a,Herrmann2010b,Herrmann2011}. These hybrid approaches have shown quite success in predicting atomization even in complex aircraft engine injectors (\cite{Kim2014,Li2013}), yet they currently have major shortcomings. Concerning the secondary atomization particularly, the effect of gas volume displacement due to the attention of droplets in the dense and intermediate regions is neglected and still lacking in the current standard two-way coupling Point-Particle approaches. Thus, to enhance and improve the current Lagrangian PP formulations as accurate predictive tools, accounting for local variations in $\theta_p$  is intended in this work. Therefore, the point-particle approach is modified by accounting for the volumetric displacements of the carried phase due to the motion of particles or droplets. The disperse phase also affects the carrier phase through mass, momentum, and energy coupling. The combined effect is termed as `volumetric coupling' which  is based on the the original formulation by \cite{Dukowicz1980} and later modified by \cite{Joseph1990}.  The approach is derived based on mixture theory that accounts for the droplet (or particle) volume fraction in a given computational cell. This effect is important in dense spray regimes, however, are typically ignored in the context of LES or DNS simulations. 

Recently, \cite{Apte2008} have shown the effect of volumetric displacements on the carrier fluid in dense particle-laden flows. They compared the results for the carrier phase and the particle dispersion obtained from the point-particle assumption and accounting for volumetric displacement effects to show large differences. If the volume displaced by the disperse phase is taken into account, the velocity field is no longer divergence free in the regions of variations in volume fractions. This has a direct effect on the pressure Poisson equation, altering the pressure field through a local source term. These effects may become important in dense regions of spray system. It was shown in the numerical work of \cite{Cihonski2013} that taking into account the volumetric displacement of fluid even under dilute loading (e.g. a small number of bubbles entrained in a vortex ring) can significantly alter the vortex core for certain combinations of the vortex strengths and bubble sizes. However, computing dense spray systems by accounting for volume displacements due to droplet motion could be numerically challenging. The temporal and spatial variations in fluid volume fractions could be locally large and make the computation numerically unstable. This is specifically true if the inter-phase coupling of mass, momentum, and energy is treated explicitly. In the present work, we focus on non-reacting flows and only momentum exchange between the two-phases is considered. As a step to spray modeling, a particle-laden turbulent jet flow similar ot the secondary atomization region will be carried out using Direct Numerical Simulation (DNS) coupled with Point-Particle/Parcel approach (PP) with quantification of the volumetric displacement effect of droplets on the flow. 

\section{Methodology}
\label{methodology}
An Eulerian--Lagrangian approach to simulate the finite-size particle--laden turbulent flows is performed in this work. Eulerian framework is used to solve the fluid equations while particles are modeled based on the Point--Particle approach using Lagrangian framework. Equations of both carrier and dispersed phases are expressed below. 

\subsection{Dispersed phase modeling}
In the Point-Particle method (e.g. \cite{Maxey1987, Elghobashi1991, Squires1991}), motion of small particles in turbulent flow field is described by a complicated integro--differential equation of \cite{Maxey1983}. Unlike body-fitted approaches (e.g. \cite{Pakseresht_2012}), the no-slip condition on the surface of particles is not imposed in this approach. The positions and velocities of individual particles are obtained by solving (\ref{newton}).

\begin{equation}
\begin{split}
\frac{d}{dt}(\mathbf{x}_p) &= \mathit{\mathbf{u}_p} \\ 
\frac{d}{dt}(\mathbf{u}_p) &= \frac{1}{m}_p\Sigma (\mathbf{F}_p) \\ 
\frac{d}{dt}(\mathbf{\Omega}_p) &= \frac{1}{i_p}\Sigma(\mathbf{T}_p) 
\label{newton} 
\end{split}
\end{equation}

\noindent Where $m_p$, $i_p$, $\mathbf{x}_p$, $\mathbf{u}_p$ and $\mathbf{\Omega}_p$ are the mass, moment of inertia, position, translational velocity and rotational velocity of each individual particle respectively. On the other hand, $\mathbf{F}_p$ and $\mathbf{T}_p$ represent deterministic forces and torques respectively acting on each particle including, closure drag by \cite{Tenneti2011}, shear induced lift force by \cite{Saffman1965}, Magnus effect of \cite{Rubinow1961}, added mass, pressure gradient and buoyancy. The effect of history force is neglected in this work as it was observed by \cite{Bagchi2003} to be insignificant. In addition to the forces, rotation and torques of each particle are performed through the closure of hydrodynamic torque given in (\ref{torques}).

\begin{equation}
\mathbf{F}_p = \mathbf{F}_g  + \mathbf{F}_{pr} + \mathbf{F}_d + \mathbf{F}_{l,Saff} + \mathbf{F}_{l,Mag} + \mathbf{F}_{am} 
\label{forces} 
\end{equation}

\begin{equation}
\mathbf{T}_p = \mathbf{T}_h  
\label{torques} 
\end{equation}

\subsection{Hydrodynamic forces and torques}
Hydrodynamic forces acting on each individual particle employ closures developed by theory, experiment or fully resolved simulations. Here, closure of all hydrodynamic forces and torques ($T_h$) as well as gravity and pressure forces are given through (\ref{gravity})-(\ref{T_h}). 

\begin{equation}
\mathbf{F}_g = -m_p\mathbf{g} 
\label{gravity} 
\end{equation}

\begin{equation}
\mathbf{F}_{pr} = -V_p \nabla P_{|p}
\label{pressure} 
\end{equation}

\begin{equation}
   \mathbf{F}_d  = m_p \frac{C_d(Re_p,\Theta_p)}{\tau_p}(\mathbf{u}_{f|p} -\mathbf{u}_p)    
\label{drag} 
\end{equation}

\begin{equation}
\begin{split}
\mathbf{F}_{l,Saff} &= m_pC_l \frac{\rho_f}{\rho_p}(\mathbf{u}_{f|p} -\mathbf{u}_p)\times (\nabla \times \mathbf{u}_f)_{|p}, \\
  C_l &= \frac{1.61\times6}{\pi d_p}\sqrt{\frac{\mu_f}{\rho_f}|(\nabla\times \mathbf{u}_f)_{|p}|}
\label{saffman} 
\end{split}
\end{equation}

\begin{equation}
\begin{split}
\mathbf{F}_{am} = m_pC_{am} \frac{\rho_f}{\rho_p}(\frac{D\mathbf{u}_{f|p}}{Dt} - \frac{d\mathbf{u}_p}{dt}), \quad C_{am} = 0.5
\label{addmass} 
\end{split}
\end{equation}

\begin{equation}
\mathbf{F}_{l,Mag} = C_{mag}\frac{\mathbf{u}_{rel} \times \mathbf{\Omega}_{rel}}{|\mathbf{\Omega}_{rel}|}(\frac{1}{2}\rho_f |\mathbf{u}_{rel}|A), \quad \text{and} \quad C_{mag} = min(0.5,0.25\frac{d_p |\omega_{rel}|}{|u_{rel}|})
\label{magnus} 
\end{equation}

\begin{equation}
\mathbf{T}_h = i_p\frac{60}{64\pi}\frac{\rho_f}{\rho_p}C_t|\mathbf{\Omega}_{rel}|\mathbf{\Omega}_{rel}.  
\label{T_h} 
\end{equation}

\noindent where volume and volume fraction of each particle are represented by $V_p$ and $\Theta_p$ respectively.
$\mathbf{u}_{rel}= \mathbf{u}_{f|p}-\mathbf{u}_p$ is the relative velocity between fluid velocity seen by particle ($\mathbf{u}_{f|p}$) and particle velocity ($\mathbf{u}_p$) so is the relative rotational velocity ($\mathbf{\Omega}_{rel}$). On the other hand, $\tau_p=(\rho_p{d_p}^2)/(18\rho_f \mu_f \Theta_f)$ and $Re_p=(\Theta_{f|p} \rho_f |u_{rel}|d_p)/(\mu_f)$ are the respective particle relaxation time and Reynolds number modified with the volumetric displacement effect \cite{Finn2016}. $C_t$ in \ref{T_h} is determined based on work of \cite{Pan2001}. Due to high loading of droplets in the dense sprays, modified coefficient of drag based on work of \cite{Tenneti2011} given in \ref{cd} is employed here.

\begin{equation}
\centering 
\begin{split}
&C_d(Re_p,\Theta_p) = (1-\Theta_p)(\frac{C_d(Re_p,0)}{(1-\Theta_p)^3} + A + B),\\
&A = \frac{5.81\Theta_p}{(1-\Theta_p)^3}+0.48\frac{\Theta^{1/3}}{(1-\Theta_p)^4}, \\
&B = \Theta^3_p Re_p(0.95+\frac{0.61\Theta^3_p}{(1-\Theta_p)^2}), \\
&C_d(Re_p,0) = 1 + 0.15 Re^{0.687}_p.
\label{cd} 
\end{split}     
\end{equation}

\subsection{Fluid Phase Modeling}
Direct Numerical Simulation (DNS) is used to solve the fluid equations in structured Cartesian grid using finite volume discretization. A pressure based second order fractional time step method based on work of \cite{Cihonski2013}, \cite{Finn2011} and \cite{Shams2011} adjusted to co-located structured grid by \cite{Finn2016} is used here. To consider the effect of fluid volume displaced by the motion of particles, the volume filtered Navier-Stokes equations given in (\ref{mass})-(\ref{moment}) are applied here \cite{Anderson1967}.

\begin{equation}
\frac{\partial (\rho_f \theta_f)}{\partial t} + \nabla \cdot (\rho_f \theta_f\mathbf{u}_f) = 0.
\label{mass}
\end{equation}

\noindent where $\rho_f$, $\theta_f$, and ${\bf{u}}_f$ are density, concentration, and velocity of the fluid phase respectively. Fluid concentration is calculated as $\theta_f = 1 - \theta_p$, 
where $\theta_p$ is particle concentration. Local spatio-temporal variations of particle concentration, generate a non-divergence free velocity field in the flow \citep{pakseresht_2017_aps}. Modified momentum equations are also given in (\ref{moment}). 

\begin{multline} 
\frac{\partial (\rho_f \theta_f \mathbf{u_f} ) }{\partial t} + \nabla \cdot ( \rho_f \theta_f \mathbf{u_f} \mathbf{u_f}) =  \\
-\nabla P + \nabla \cdot (\mu_f(\nabla \mathbf{u}_f + \nabla \mathbf{u}^T_f - \frac{2}{3} \nabla \cdot \mathbf{u}_f) ) \\ 
- \theta_f \rho_f \mathbf{g} + \mathbf{F}_{p \rightarrow f} + \nabla \cdot \mathbf{T}_{p \rightarrow f}.
\label{moment}
\end{multline}

\noindent In the above form, the conservation equations (i.e. both continuity and momentum equations) account for the volume of fluid displaced by the motion of particles through the fluid volume fraction of $\theta_f$. In addition, (\ref{moment}) also contains the typical inter-phase momentum transfer term (two-way coupling), $\mathbf{F}_{p \rightarrow f}$, based on particle forces. Moreover, to accurately capturing the effect of particles onto the flow, additional inter-phase momentum transfer through the rotation of particles, $\nabla \cdot \mathbf{T}_{p \rightarrow f}$ based on work of \cite{Andersson2012} is added to the momentum equations (referred to as Torque coupling). These terms, $\mathbf{F}_{p \rightarrow f}$ and $\nabla \cdot \mathbf{T}_{p \rightarrow f}$, include the equal and opposite reaction from the particle surface forces and torques respectively back to the flow. Accordingly, it is crucial to define a function, $f(\mathbf{x},\mathbf{x}_p)$ given in (\ref{projection}) to project Lagrangian quantities of particles, $\psi_p$, back to the continuous field, $\psi_f$, located at Eulerian grid points. Likewise, fluid properties are required to be interpolated to the particles' position as well. Gaussian function for both interpolation and projection purposes is employed with bandwidth of local grid size around each particle. It is worth mentioning that giving $\theta_f=1$ switches the above formulation to the standard two-way coupling where effect of the volume displacement of the carrier phase is neglected. 
 
\begin{equation}
\psi_f(\mathbf{x}) = \sum_{ip=1}^{n_p} {f(\mathbf{x},\mathbf{x}_p)\psi_p} 
\label{projection}
\end{equation}

\section{Results and Discussion}
\label{sec:result}
The above numerical scheme is applied to different test cases in order to evaluate its accuracy and robustness (e.g. \cite{Pakseresht_2014,Pakseresht_2015,Pakseresht2016,Pakseresht_2017_ASME}). One of these test cases are described below and will be followed by particle-laden turbulent jet as a step in spray modeling. 

\subsection{Oscillating bubble}
First the importance of volumetric displacement effect on the flow filed caused by change in local concentration of particles is verified in this section. The variable density formulation used here accounts for changes in the density of mixture. Here, a very simple case of imposed oscillation on the radius of a bubble which causes a potential flow field around itself is set up. This phenomenon can not be simulated by only inter-phase momentum coupling and here it is observed that only through the variations in density in momentum and continuity equation, the potential flow is expressed.

\begin{figure}[ht]
\centering
\includegraphics[scale=0.35]{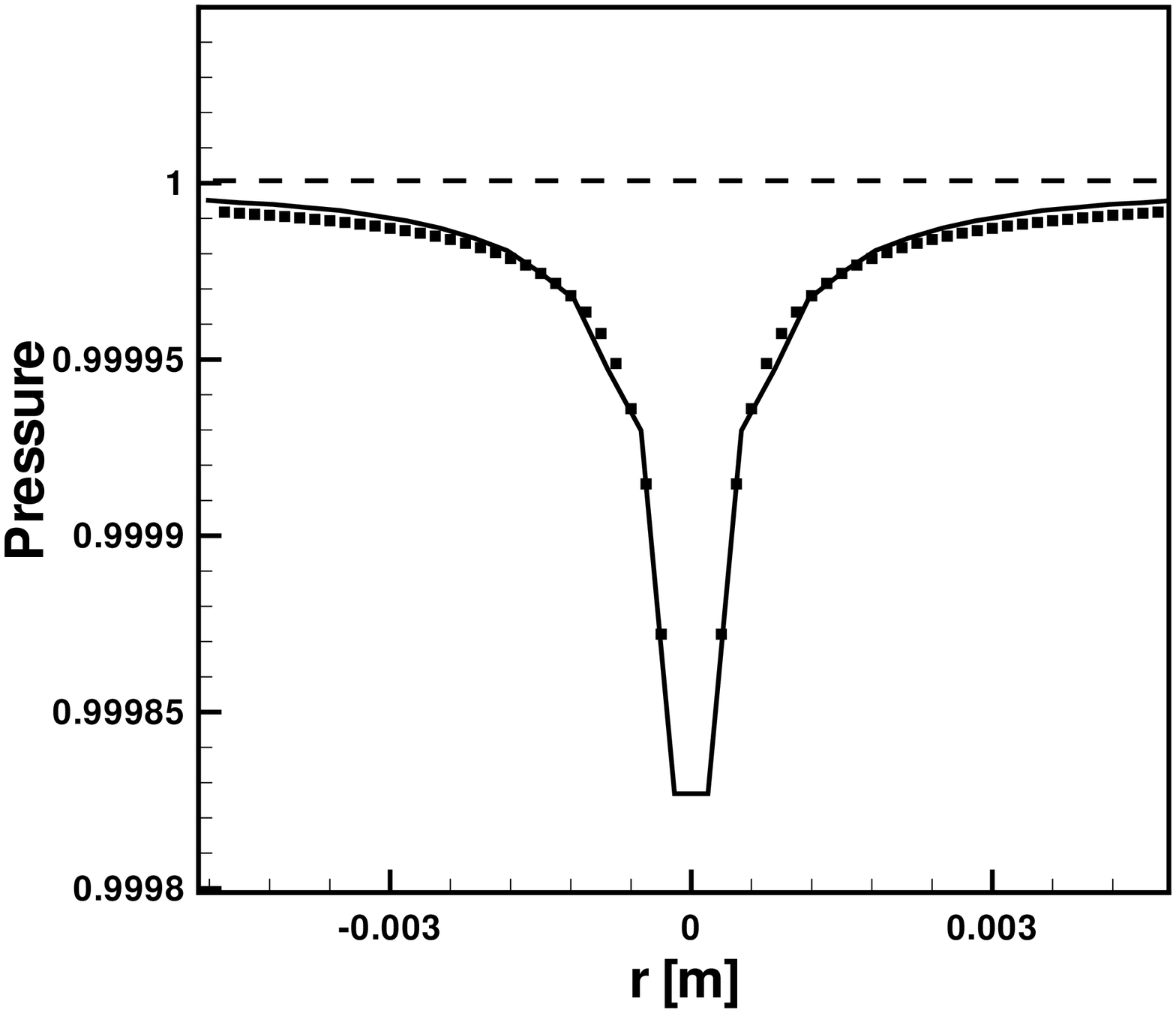}
\caption{Pressure distribution caused by volume displacement around the bubble, from two-way coupling (dashed line), volumetric coupling (solid line), and analytical solution (dots).}
\label{fig:p_single}
\end{figure}

A single air bubble in a cube of water is located and sinusoidal perturbation on the bubble radius is imposed. Bubble radius changes in time as $R=R_0 + e \sin(\omega t)$, where $R$ and $R_0$ are the instantaneous and the initial radius, respectively, $e$ is the perturbation magnitude, $\omega$ is frequency and $t$ is time. In this simulation, $R_0 = 0.01 \times D$, where $D$ is the cube size, and gives overall concentration of $4\times 10^{-6}$, $e=0.1 \times R_0$, $\omega=50 [Hz]$. Figure~\ref{fig:p_single} shows the radial distribution of hydrodynamic pressure around the bubble created by the size variation at $t^* = 0.3$ where $t^* = t / T$ and $T=2\pi/\omega$. Analytical solution for pressure (dots), given by~\cite{Panton2006} is compared and shown in Fig.\ref{fig:p_single} with result of two-way coupling and no volumetric effect (dashed line). No any effect in the result of two-way coupling for pressure is achieved, however the volumetric coupling result (solid line) is in good agreement with the analytical solution \citep{Shams2011}. 

\begin{figure}[ht]
\centering
\includegraphics[scale=0.35]{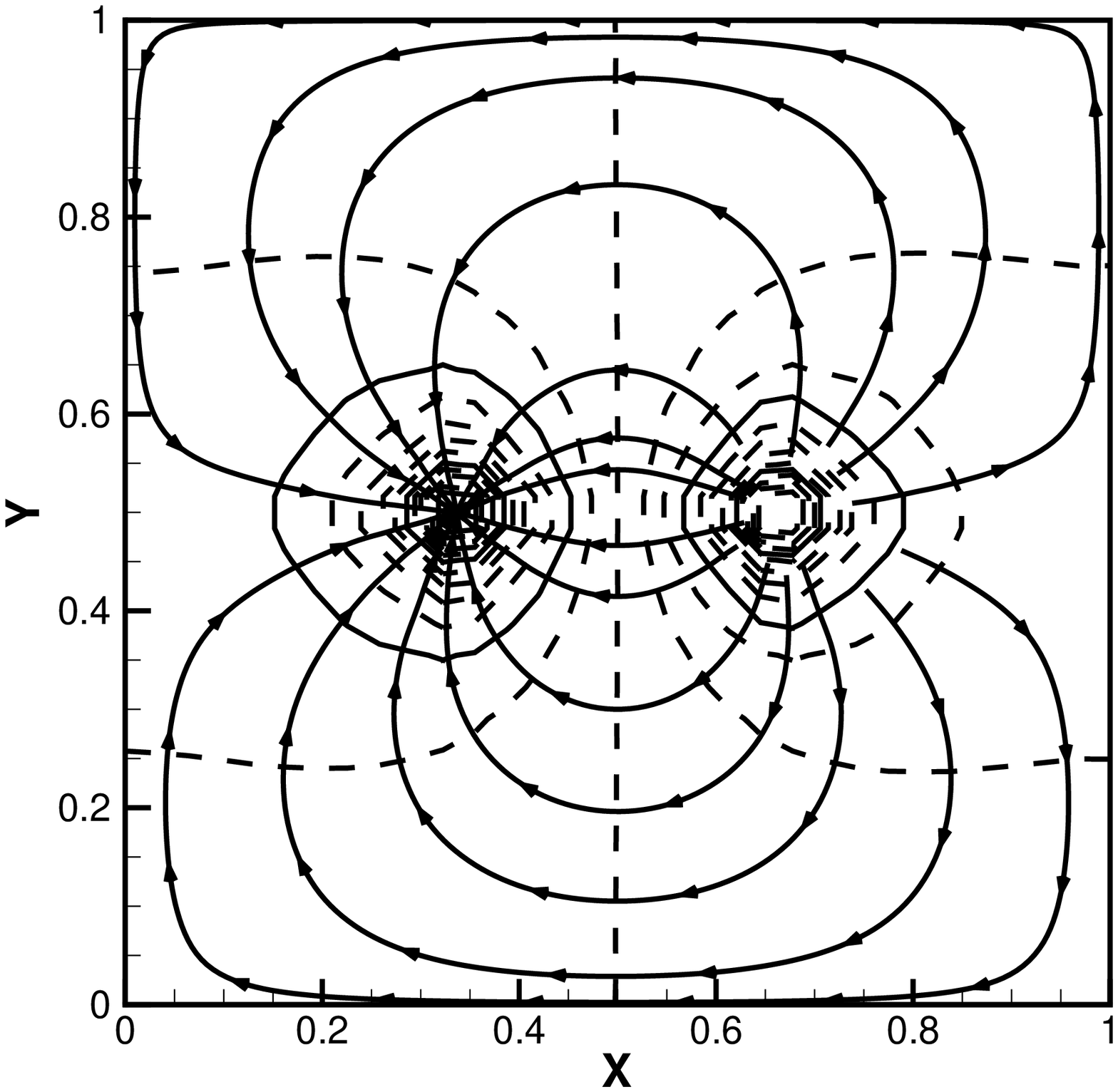}
\caption{Doublet generated by bubbles oscillating in tandem}
\label{fig:doublet}
\end{figure}

In another similar example, two bubbles oscillating in tandem are considered. Two similar bubbles are put in a box and their radius changes sinusoidally with $\pi$ [rad] phase shift. All properties are similar to the case of single bubble case, except they are both located $D/6$ away from the box center. The result is a doublet-like flow which is shown in Fig.~\ref{fig:doublet}. Likewise, in this case no any effect on the flow is observed by two-way coupling results \citep{Shams2011}. 

\subsection{Particle-laden turbulent flow}
\begin{figure}
\setlength{\unitlength}{0.012500in}%
\centering
\includegraphics[scale=0.4]{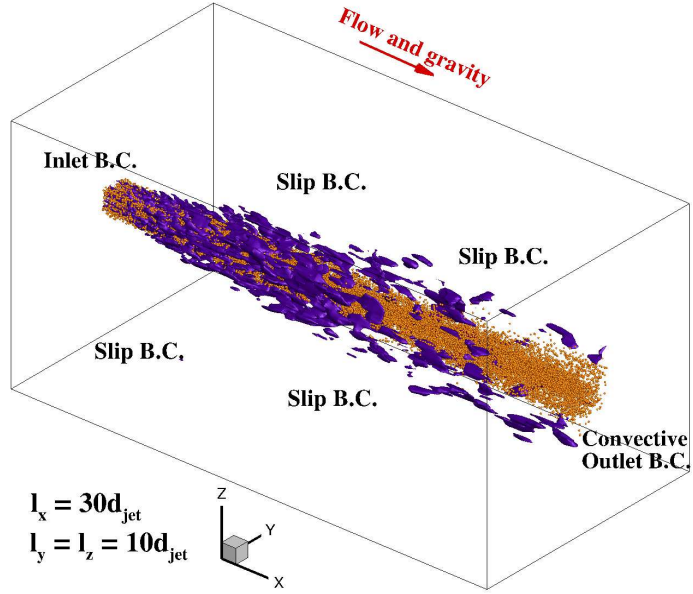}
\caption{Schematic of particle laden jet with specifying boundary conditions. Vorticity magnitude and particle distribution with exaggerating in particle size for sake of clarity are shown. }
\label{fig:schematic_jet}
\end{figure}

As a step in developing spray modeling, particle-laden turbulent jets have been studied recently (e.g. \cite{Shuen1983}) to obtain the underlying structures of suspended particles and turbulence which could eventually help understand the physics of liquid breakup in the secondary atomization process. Accordingly, in this part, a particle-laden turbulent jet flow performed by \cite{Mostafa1989} is carried out numerically using DNS along with volumetric coupling of Point-Particle tracking approach as a first step in modeling dense sprays. Flow parameters are tabulated in Tab.~\ref{tab:jet_parameters}. A Cartesian structured grid is applied for solving the flow in a rectangular computational domain. It is well recognized that properly chosen boundary conditions is of importance in view of a good representation of a physical jet. Inflow data over several flow through times is generated a priori and read at each flow time step to specify the fluid velocity components at the inlet. Convective outflow boundary condition is applied at the exit section while slip boundary condition is enforced for other sides of the computational domain as shown in Fig.~\ref{fig:schematic_jet}. To overcome on the reflection of boundary error from downstream side to the upstream side of the jet due to the hyperbolic characteristic of the convective outlet boundary condition \cite{Dai1994}, a long enough computational domain ($30d_j\times10d_j\times10d_j$) is used to obtain more accurate results yet more expensive in terms of computational cost. 

\begin{table}
\begin{center}
\begin{tabular}{|r|r|r|}\hline
Property & \multicolumn{1}{|c}{Value} & \multicolumn{1}{|c|}{Unit} \\ \hline
$d_{jet}$ & 0.0253 & $m$ \\
$\mu_f$ & $1.8502\times10^{-5}$ & $(N.s)m^{-2}$ \\
$\rho_f$ & 1.178 & $kg/m^3$ \\
$U_{jet}$ & 3.546 & $ms^{-1}$ \\
$Re_{jet}$ & 5712 & -- \\
$d_p$ & $105\times10^{-6}$ & $m$ \\
$\rho_p$ & 2500 &  $kg/m^3$ \\
$\alpha$ & 1 & --\\ \hline
\end{tabular}
\end{center}
\caption{Particle-laden jet parameters}
\label{tab:jet_parameters}
\end{table}

As mentioned earlier, inflow data is only used for the fluid phase of the flow; however, for solid phase a measured data from experiment is prescribed at the nozzle exit. It is worth mentioning that since no measured data is available right at the nozzle exit, rather at $x/d_{jet}=0.04$, particles are injected at the nozzle exit based on the provided measured mean and rms fluctuating velocities at $x/d_{jet}=0.04$ (1 mm from the nozzle exit). Concerning the number of particles, they are injected randomly through the cross section of the jet inlet based on the initial solid-air mass flux rate given in Tab.~\ref{tab:jet_parameters}. Total number of particles ($n_p$) required for injecting per flow time step ($\Delta t_f$) is calculated through Eq.~\ref{np}. It should be emphasized that injection is performed once the particle-free jet (clear jet) has reached to the statistically stable condition. To interpolate gas properties at the position of the solid phase, Gaussian function is employed with filter size of equal to the local grid resolution ($\sigma \sim \Delta$). Similarly, this function with equal filter size is used for projecting the particles' forces back onto the Eulerian framework of the gas phase.  

\begin{equation}
n_p = \frac{6\alpha \rho_f {d_j}^2u_j \Delta t_f}{4 \rho_p {d_p}^3}
\label{np}
\end{equation}

\begin{figure}[ht]
\setlength{\unitlength}{0.012500in}
\centering
\includegraphics[scale=0.5]{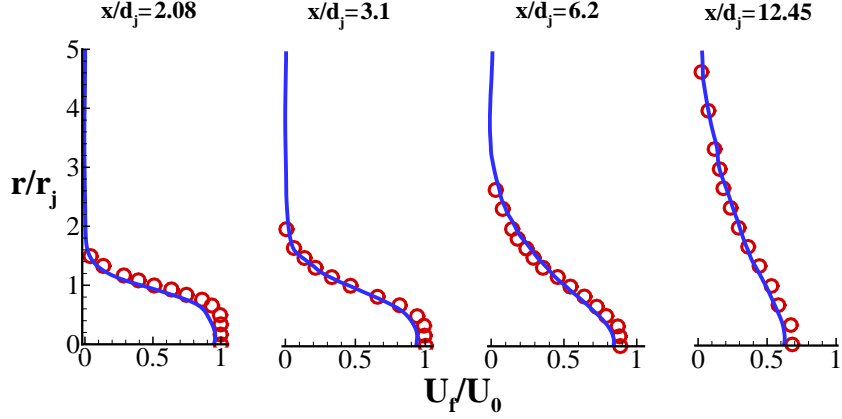}
\caption{Radial profile of normalized mean axial velocity of the gas phase, solid line: simulation, open symbol: experiment}
\label{fig:uf_jet}
\end{figure}

Simulations have been performed in the developing ($x/d_j<6$) and self-similar regions of the two-phase jet close to the nozzle. The results are plotted in a dimensionless form versus normalized radius ($r/r_j$) to show the spreading of the jet in the radial direction. All quantities except mean axial velocity of the fluid phase are normalized by the local mean centerline velocity ($U_c$) while mean axial gas velocity is normalized by the initial jet centerline velocity at the nozzle exit ($U_0$). In this way, the jet centerline velocity decay can be illustrated as well. Figure \ref{fig:uf_jet} shows the radial profile of normalized mean axial velocity of the gas phase at the several nozzle distances. As depicted in this figure, mean velocity of the gas phase at different distances from the jet exit is well predicted compared to the experiment. In addition, the decaying and spreading features of the jet can be easily observed in this plot. By getting farther from the nozzle exit, mean velocity of the jet is decayed while the jet is spread more in radial direction.

\begin{figure}[ht]
\setlength{\unitlength}{0.012500in}
\centering
\includegraphics[scale=0.5]{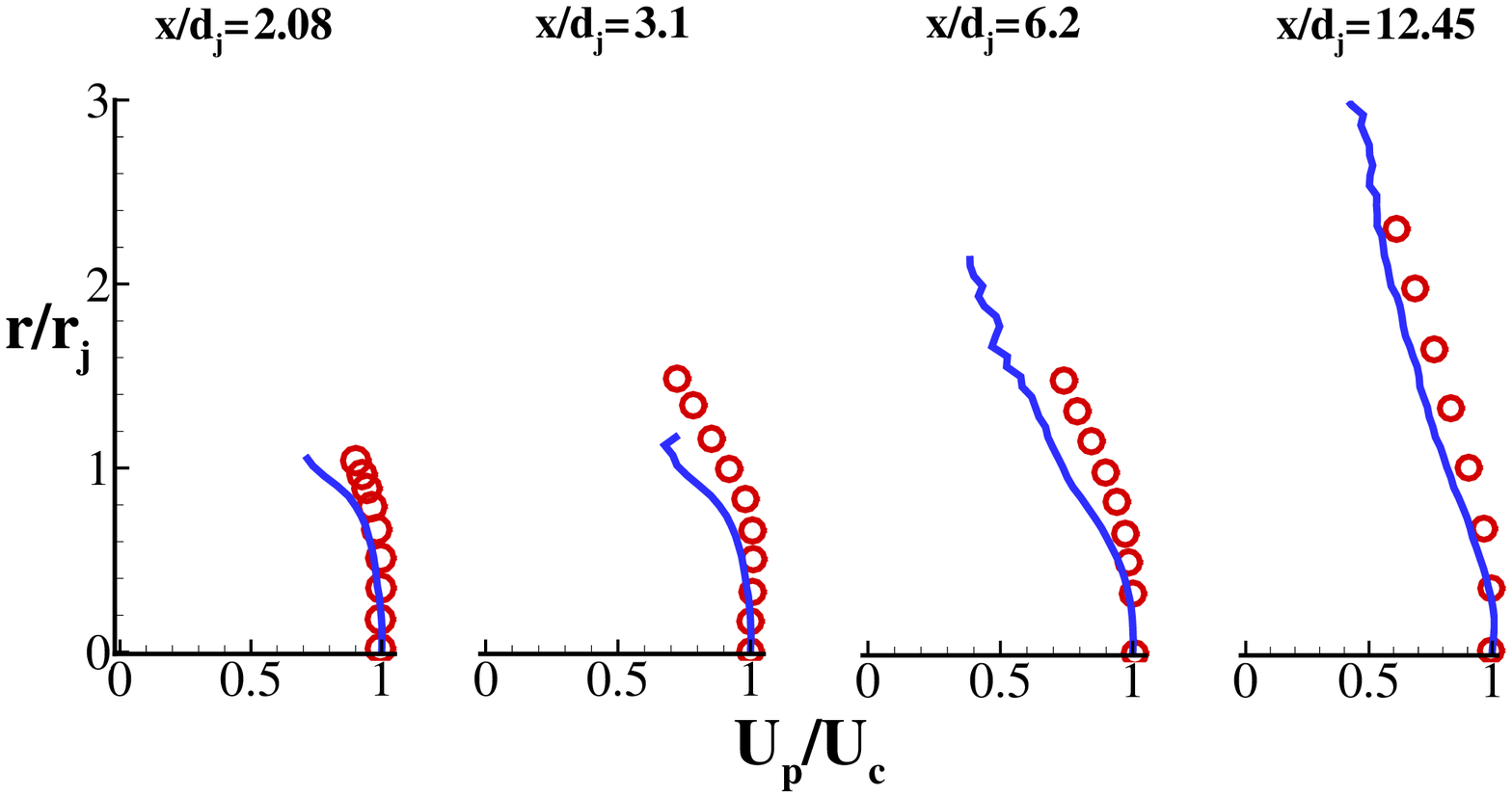}
\caption{Radial profile of normalized mean axial velocity of the solid phase, solid line: simulation, open symbol: experiment}
\label{fig:up_jet}
\end{figure}

Likewise, radial profile of the normalized mean axial velocity of the solid phase depicted in Fig.~\ref{fig:up_jet} illustrates a good agreement between simulation and the experiment; however, slightly under prediction in the edge of the jet is noticeable in the simulation. It should be emphasized that similar results for the solid phase were obtained in the numerical prediction part of \cite{Mostafa1989} (not shown here). This discrepancy between numerical simulation and experiment for solid phase can be addressed as a result of inappropriate inflow data for solid phase. Particles would have been more accurately predicted if they had been generated as priori in line with gas phase. As a result, a remedy to overcome this discrepancy would be generating the inflow data in which particles have been already injected and interacted with the carrier phase. This would definitely affect the carrier phase and give rise to accurately improve the simulation results. 

\begin{figure}[ht]
\setlength{\unitlength}{0.012500in}%
\centering
\includegraphics[scale=0.5]{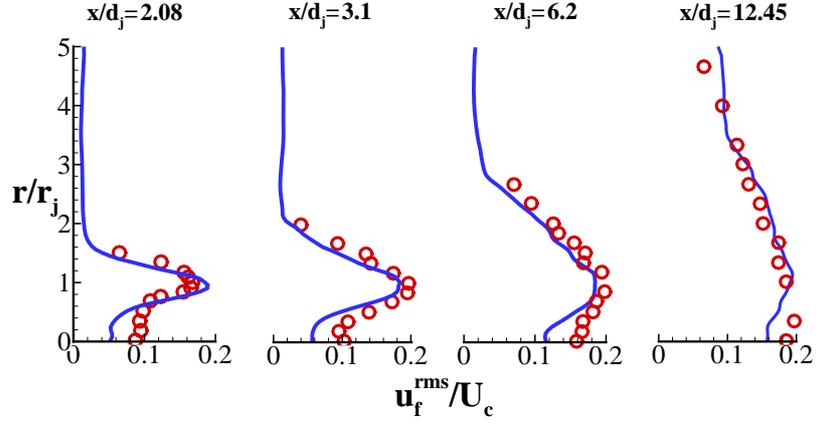}
\caption{Radial profile of normalized axial fluctuating velocity of gas phase, solid line: simulation, open symbol: experiment}
\label{fig:rmsf_jet}
\end{figure}

 \begin{figure}[ht]
\setlength{\unitlength}{0.012500in}%
\centering
\includegraphics[scale=0.4]{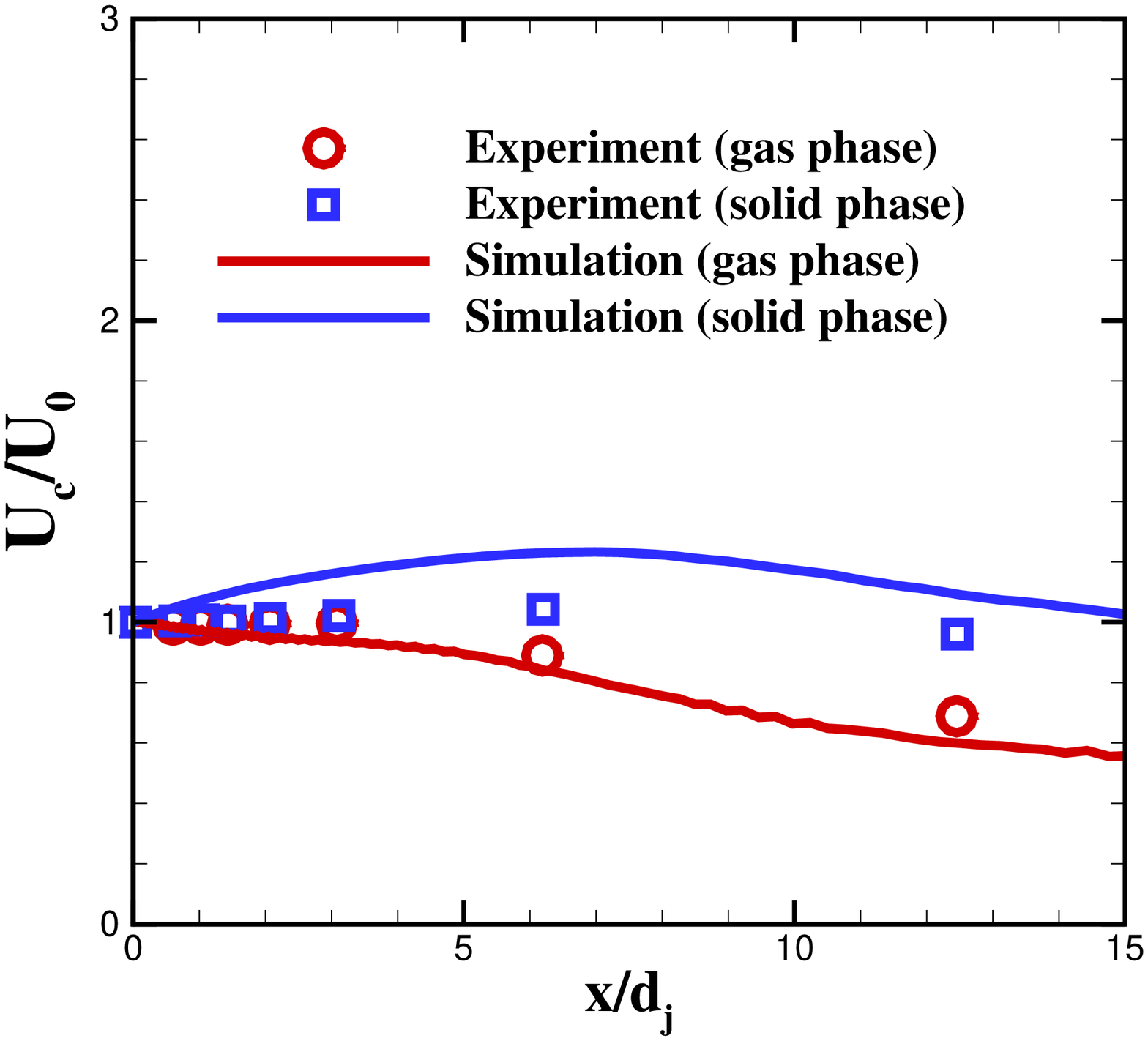}
\caption{Centerline velocity for both gas and solid phases}
\label{fig:center_vel}
\end{figure}

Further insight can be obtained by looking at the dynamics of the particle laden-jet flow through the calculation of rms of fluctuating axial velocity for fluid phase. As shown in Fig.~\ref{fig:rmsf_jet}, for the region of $x/d_j>2.08$, it can be clearly seen that simulation for rms of gas phase is within the range of experiment with slightly lower values in $r/r_j>0.5$ as well as significant under prediction in the radial region of $r/r_j<0.5$ compared to the experiment. This reveals the fact that the grid resolution is not sufficient to capture all the smallest scales embedded in the gas phase. Higher resolution would definitely help improve the rms of fluid particularly in the shear region of the jet where high gradient of velocity exists. 

Finally, the axial profile of the predicted mean velocity in the center of the jet for both phases compared to the experiment are plotted in Fig.\ref{fig:center_vel}. The velocity profile for each phase is normalized by its corresponding mean velocity in the center of the nozzle exit (i.e. $U_{f0}$ for gas phase and $U_{p0}$ for solid phase). As depicted in this figure, both phases are well predicted in the center line mean velocity; however, slightly higher velocity for solid phase while lower velocity for gas phase is noticeable in the simulation. 

\section{Summary and conclusion}
A numerical formulation based on co-located grid, finite volume approach is developed for simulation of dense particle-laden flows. Direct Numerical Simulation coupled with point-particle method was performed to model the interactions of particle and turbulence in the particle-laden jet flows as well as secondary atomization region in sprays. Volumetric displacement effect of fluid due to presence of liquid particles was taken into account to capture accurately the underlying structure of particle-laden turbulent flows. Several test cases were considered to evaluate the accuracy and robustness of the numerical scheme for dense loadings. First, the effect of a single bubble undergoing forced periodic oscillations was computed by considering the present approach as well as the standard `two-way' coupling based point-particle method to show large variations in the predicted flow field. The results were compared with analytical solutions to validate the numerical approach for volumetric coupling. A test case with two bubbles undergoing forced oscillations in tandem was also investigated. The doublet-like flow pattern was well predicted by the present approach. Next, a particle-laden turbulent round jet with $Re=5712$ based on work of \cite{Mostafa1989} was simulated to test the robustness of the numerical scheme with taking into account the volumetric displacement of fluid due to presence of particles. The present numerical approach is capable to simulate the dense regime of particle-laden turbulent jet flows or sprays. 

\section{Nomenclature}
\begin{tabular}{ll}
$d_{jet}$     & Jet diameter \\
$d_p$ & Particle diameter \\
$Re_{jet}$    & Jet Reynolds number \\
$U_{jet}$     & Jet bulk velocity \\
$\rho_p$ & Particle density\\
$\alpha$ & Mass loading\\
$\mu_f$ & Air viscosity \\
$\rho_f$ & Air density \\
$St$ & Stokes number\\
$\tau_p$ & Particle relaxation time\\
$m_p$ & Particle mass\\
$i_p$ & Particle inertia\\
$\mathbf{u}_p$ & Particle translational velocity vector\\
$\mathbf{\Omega}_p$ & particle rotational velocity vector\\
$\mathbf{F}_p$ & Deterministic particle force\\
$\mathbf{T}_p$ & Deterministic particle torque\\
$V_p$ & Particle volume\\
$\theta_p$ & Volume fraction of particle\\
$\mathbf{u}_{rel}$ & Relative velocity between phases \\
$\mathbf{u}_{f|p}$ & Fluid velocity seen by particle\\
& \\
\end{tabular}

\bibliographystyle{elsarticle-harv}\biboptions{authoryear}
\bibliography{manuscript}

\begin{thebibliography}{37}
\expandafter\ifx\csname natexlab\endcsname\relax\def\natexlab#1{#1}\fi
\expandafter\ifx\csname url\endcsname\relax
  \def\url#1{\texttt{#1}}\fi
\expandafter\ifx\csname urlprefix\endcsname\relax\def\urlprefix{URL }\fi

\bibitem[{Anderson and Jackson(1967)}]{Anderson1967}
Anderson, T.~B., Jackson, R., 1967. Fluid mechanical description of fluidized
  beds. equations of motion. Industrial \& Engineering Chemistry Fundamentals
  6~(4), 527--539.

\bibitem[{Andersson et~al.(2012)Andersson, Zhao, and Barri}]{Andersson2012}
Andersson, H.~I., Zhao, L., Barri, M., 2012. Torque-coupling and
  particle--turbulence interactions. Journal of Fluid Mechanics 696, 319--329.

\bibitem[{Apte et~al.(2008)Apte, Mahesh, and Lundgren}]{Apte2008}
Apte, S., Mahesh, K., Lundgren, T., 2008. Accounting for finite-size effects in
  simulations of disperse particle-laden flows. International Journal of
  Multiphase Flow 34~(3), 260--271.

\bibitem[{Bagchi and Balachandar(2003)}]{Bagchi2003}
Bagchi, P., Balachandar, S., 2003. Effect of turbulence on the drag and lift of
  a particle. Physics of fluids 15~(11), 3496--3513.

\bibitem[{Blokkeel et~al.(2003)Blokkeel, Barbeau, and Borghi}]{Blokkeel2003}
Blokkeel, G., Barbeau, B., Borghi, R., 2003. A 3d eulerian model to improve the
  primary breakup of atomizing jet. Tech. rep., SAE Technical Paper.

\bibitem[{Chesnel et~al.(2011{\natexlab{a}})Chesnel, Menard, Reveillon, and
  Demoulin}]{Chesnel2011b}
Chesnel, J., Menard, T., Reveillon, J., Demoulin, F.-X., 2011{\natexlab{a}}.
  Subgrid analysis of liquid jet atomization. Atomization and Sprays 21~(1).

\bibitem[{Chesnel et~al.(2011{\natexlab{b}})Chesnel, Reveillon, Menard, and
  Demoulin}]{Chesnel2011a}
Chesnel, J., Reveillon, J., Menard, T., Demoulin, F.-X., 2011{\natexlab{b}}.
  Large eddy simulation of liquid jet atomization. Atomization and Sprays
  21~(9).

\bibitem[{Cihonski et~al.(2013)Cihonski, Finn, and Apte}]{Cihonski2013}
Cihonski, A.~J., Finn, J.~R., Apte, S.~V., 2013. Volume displacement effects
  during bubble entrainment in a travelling vortex ring. Journal of Fluid
  Mechanics 721, 225--267.

\bibitem[{Dai et~al.(1994)Dai, Kobayashi, and Taniguchi}]{Dai1994}
Dai, Y., Kobayashi, T., Taniguchi, N., 1994. Large eddy simulation of plane
  turbulent jet flow using a new outflow velocity boundary condition. JSME
  International Journal Series B Fluids and Thermal Engineering 37~(2),
  242--253.

\bibitem[{Dukowicz(1980)}]{Dukowicz1980}
Dukowicz, J.~K., 1980. A particle-fluid numerical model for liquid sprays.
  Journal of Computational Physics 35~(2), 229--253.

\bibitem[{Elghobashi(1991)}]{Elghobashi1991}
Elghobashi, S., 1991. Particle-laden turbulent flows: direct simulation and
  closure models. Applied Scientific Research 48~(3-4), 301--314.

\bibitem[{Finn et~al.(2011)Finn, Shams, and Apte}]{Finn2011}
Finn, J., Shams, E., Apte, S.~V., 2011. Modeling and simulation of multiple
  bubble entrainment and interactions with two dimensional vortical flows.
  Physics of Fluids 23~(2), 023301.

\bibitem[{Finn et~al.(2016)Finn, Li, and Apte}]{Finn2016}
Finn, J.~R., Li, M., Apte, S.~V., 2016. Particle based modelling and simulation
  of natural sand dynamics in the wave bottom boundary layer. Journal of Fluid
  Mechanics 796, 340--385.

\bibitem[{Herrmann(2010{\natexlab{a}})}]{Herrmann2010b}
Herrmann, M., 2010{\natexlab{a}}. Detailed numerical simulations of the primary
  atomization of a turbulent liquid jet in crossflow. Journal of Engineering
  for Gas Turbines and Power 132~(6), 061506.

\bibitem[{Herrmann(2010{\natexlab{b}})}]{Herrmann2010a}
Herrmann, M., 2010{\natexlab{b}}. A parallel eulerian interface
  tracking/lagrangian point particle multi-scale coupling procedure. Journal of
  Computational Physics 229~(3), 745--759.

\bibitem[{Herrmann(2011)}]{Herrmann2011}
Herrmann, M., 2011. On simulating primary atomization using the refined level
  set grid method. Atomization and Sprays 21~(4).

\bibitem[{Joseph et~al.(1990)Joseph, Lundgren, Jackson, and
  Saville}]{Joseph1990}
Joseph, D., Lundgren, T., Jackson, R., Saville, D., 1990. Ensemble averaged and
  mixture theory equations for incompressible fluid-particle suspensions.
  International journal of multiphase flow 16~(1), 35--42.

\bibitem[{Kim et~al.(2014)Kim, Ham, Le, Herrmann, Li, Soteriou, and
  Kim}]{Kim2014}
Kim, D., Ham, F., Le, H., Herrmann, M., Li, X., Soteriou, M., Kim, W., 2014.
  High-fidelity simulation of atomization in a gas turbine injector high shear
  nozzle. In: ILASS Americas 26th Annual Conference on Liquid Atomization and
  Spray Systems, Portland, OR, May. pp. 18--21.

\bibitem[{Lebas et~al.(2005)Lebas, Blokkeel, Beau, and Demoulin}]{Lebas2005}
Lebas, R., Blokkeel, G., Beau, P.-A., Demoulin, F.-X., 2005. Coupling
  vaporization model with the eulerian-lagrangian spray atomization (elsa)
  model in diesel engine conditions. Tech. rep., SAE Technical Paper.

\bibitem[{Li and Soteriou(2013)}]{Li2013}
Li, X., Soteriou, M., 2013. High-fidelity simulation of fuel atomization in a
  realistic swirling flow injector. Atomization and Sprays 23~(11).

\bibitem[{Maxey(1987)}]{Maxey1987}
Maxey, M., 1987. The gravitational settling of aerosol particles in homogeneous
  turbulence and random flow fields. Journal of Fluid Mechanics 174, 441--465.

\bibitem[{Maxey and Riley(1983)}]{Maxey1983}
Maxey, M.~R., Riley, J.~J., 1983. Equation of motion for a small rigid sphere
  in a nonuniform flow. The Physics of Fluids 26~(4), 883--889.

\bibitem[{Mostafa et~al.(1989)Mostafa, Mongia, McDonell, and
  Samuelsen}]{Mostafa1989}
Mostafa, A., Mongia, H., McDonell, V., Samuelsen, G., 1989. Evolution of
  particle-laden jet flows-a theoretical and experimental study. AIAA journal
  27~(2), 167--183.

\bibitem[{Pakseresht et~al.(2014)Pakseresht, Apte, and Finn}]{Pakseresht_2014}
Pakseresht, P., Apte, S., Finn, J., 2014. Ineractions of turbulence and
  sediment particles in an open channel flow. In: APS Meeting Abstracts.

\bibitem[{Pakseresht et~al.(2015)Pakseresht, Apte, and Finn}]{Pakseresht_2015}
Pakseresht, P., Apte, S., Finn, J., 2015. Dns with discrete element modeling of
  suspended sediment particles in an open channel flow. In: APS Division of
  Fluid Dynamics Meeting Abstracts.

\bibitem[{Pakseresht et~al.(2016)Pakseresht, Apte, and Finn}]{Pakseresht2016}
Pakseresht, P., Apte, S., Finn, J., 2016. Dns-dem of suspended sediment
  particles in an open channel flow. In: APS Meeting Abstracts.

\bibitem[{Pakseresht and Apte(2017)}]{pakseresht_2017_aps}
Pakseresht, P., Apte, S.~V., 2017. Prediction of a densely loaded
  particle-laden jet using a euler-lagrange dense spray model. In: APS Division
  of Fluid Dynamics Meeting Abstracts.

\bibitem[{Pakseresht et~al.(2017)Pakseresht, Apte, and
  Finn}]{Pakseresht_2017_ASME}
Pakseresht, P., Apte, S.~V., Finn, J.~R., 2017. On the predictive capability of
  dns-dem applied to suspended sediment-turbulence interactions. In: ASME 2017
  Fluids Engineering Division Summer Meeting. American Society of Mechanical
  Engineers Digital Collection.

\bibitem[{Pakseresht et~al.(2012)Pakseresht, Bahrainian, and
  Bahoosh~Kazerooni}]{Pakseresht_2012}
Pakseresht, P., Bahrainian, S., Bahoosh~Kazerooni, R., 2012. An algorithm in
  element deletion process used in moving three zones unstructured grid for
  arbitrary relative motion of two objects. In: Proceedings of 3 rd
  International Conference on Theoretical and Applied Mechanics, Athens, Gr.

\bibitem[{Pan et~al.(2001)Pan, Tanaka, and Tsuji}]{Pan2001}
Pan, Y., Tanaka, T., Tsuji, Y., 2001. Direct numerical simulation of
  particle-laden rotating turbulent channel flow. Physics of Fluids 13~(8),
  2320--2337.

\bibitem[{Panton(2006)}]{Panton2006}
Panton, R., 2006. Incompressible Flow. Wiley-Interscience.

\bibitem[{Rubinow and Keller(1961)}]{Rubinow1961}
Rubinow, S., Keller, J.~B., 1961. The transverse force on a spinning sphere
  moving in a viscous fluid. Journal of Fluid Mechanics 11~(03), 447--459.

\bibitem[{Saffman(1965)}]{Saffman1965}
Saffman, P., 1965. The lift on a small sphere in a slow shear flow. Journal of
  fluid mechanics 22~(02), 385--400.

\bibitem[{Shams et~al.(2011)Shams, Finn, and Apte}]{Shams2011}
Shams, E., Finn, J., Apte, S., 2011. A numerical scheme for euler--lagrange
  simulation of bubbly flows in complex systems. International Journal for
  Numerical Methods in Fluids 67~(12), 1865--1898.

\bibitem[{Shuen et~al.(1983)Shuen, Chen, and Faeth}]{Shuen1983}
Shuen, J., Chen, L., Faeth, G., 1983. Predictions of the structure of
  turbulent, particle-laden round jets. AIAA journal 21~(11), 1483--1484.

\bibitem[{Squires and Eaton(1991)}]{Squires1991}
Squires, K.~D., Eaton, J.~K., 1991. Preferential concentration of particles by
  turbulence. Physics of Fluids A: Fluid Dynamics 3~(5), 1169--1178.

\bibitem[{Tenneti et~al.(2011)Tenneti, Garg, and Subramaniam}]{Tenneti2011}
Tenneti, S., Garg, R., Subramaniam, S., 2011. Drag law for monodisperse
  gas--solid systems using particle-resolved direct numerical simulation of
  flow past fixed assemblies of spheres. International journal of multiphase
  flow 37~(9), 1072--1092.

\end{thebibliography}

\end{document}